\documentclass[sigconf]{acmart}
\usepackage{epigraph}
\usepackage{xcolor}

\definecolor{celadon}{rgb}{0.67, 0.88, 0.69}
\definecolor{atomictangerine}{rgb}{1.0, 0.6, 0.4}
\AtBeginDocument{}

\setcopyright{none}

\copyrightyear{2024}
\acmYear{2024}

\acmPrice{15.00}
\acmISBN{978-1-4503-XXXX-X/18/06}

\begin{document}

\newcommand{\method}[0]{\texttt{CAESURA}}
\title{\method{}: Language Models as Multi-Modal Query Planners}

\author{Matthias Urban}
\affiliation{\institution{Technical University of Darmstadt}
  \city{}
  \country{}
}

\author{Carsten Binnig}
\affiliation{\institution{Technical University of Darmstadt \& DFKI}
  \city{}
  \country{}
}

\begin{abstract}
Traditional query planners translate SQL queries into query plans to be executed over relational data.
However, it is impossible to query other data modalities, such as images, text, or video stored in modern data systems such as data lakes using these query planners.
In this paper, we propose Language-Model-Driven Query Planning, a new paradigm of query planning that uses Language Models to translate natural language queries into executable query plans.
Different from relational query planners, the resulting query plans can contain complex operators that are able to process arbitrary modalities.
As part of this paper, we present a first GPT-4 based prototype called \method{} and show the general feasibility of this idea on two datasets.
Finally, we discuss several ideas to improve the query planning capabilities of today's Language Models.
\end{abstract}

\maketitle

\begin{figure}[t]
 \centering
 \includegraphics[width=0.9\linewidth]{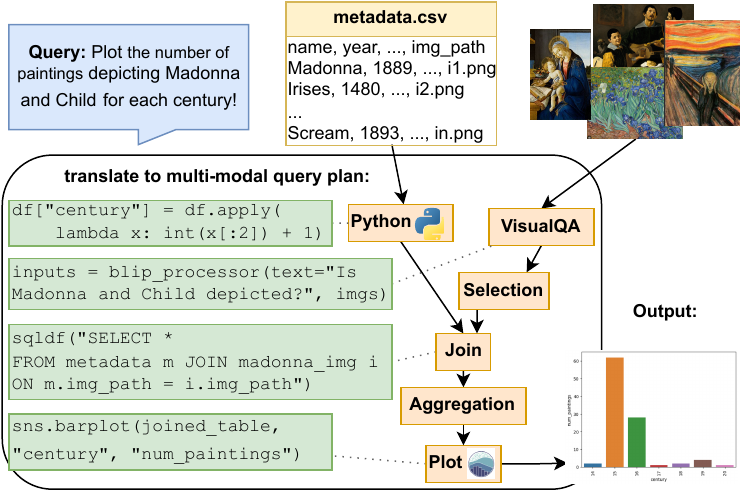}
 \vspace{-3.5ex}
 \caption{Example illustrating how a natural language query is automatically translated into a multi-modal query plan containing relational operators, machine generated Python UDFs, and a VisualQA Machine Learning model. The output is presented as a plot, making it easy and fast to gain insights from multi-modal data. The green boxes show code snippets that are executed when executing the plan.}
 \vspace{-4.5ex}
 \label{fig:teaser}
\end{figure}

\section{Introduction}

Query planning, the basic process of deriving an executable query plan in response to a user query, has conceptually stayed essentially the same since IBM's System R was introduced in 1974 \cite{selingerAccessPathSelection1979}.
In traditional DBMSs, a logical plan is first obtained from parsing a SQL query and then optimized by improving the order of the query's operators.
In a second step, each logical operator is mapped to a concrete implementation to obtain a physical plan, which is eventually executed.
In the past decades, research has focused primarily on improving the efficiency of query plans by improving various individual aspects (e.g. the cost model \cite{hilprechtZeroshotCostModels2022}).

However, this traditional approach to query planning is fundamentally limited in two important aspects:
Firstly, it only applies to query languages such as SQL, where semantics of queries are clear and can be easily parsed into a (at least canonical) query plan to execute the query.
Secondly, it is only possible to query structured relational data stored in tables.
Due to these limitations and several new trends, we argue that it is finally time to re-think how query planning is done:

\begin{figure*}[ht]
 \centering
 \includegraphics[width=0.9\linewidth]{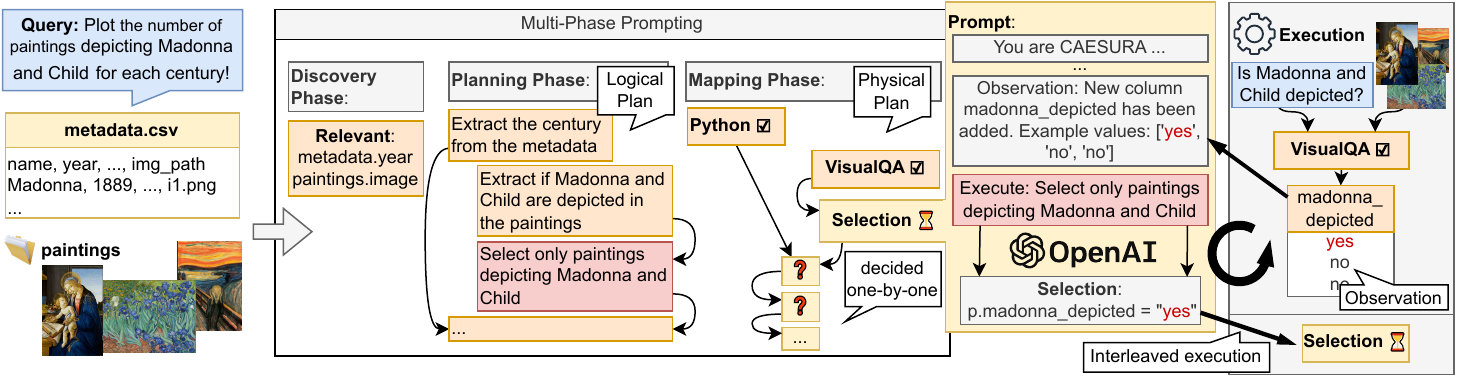}
 \vspace{-3.5ex}
 \caption{\method{} transforms the query into a multi-modal query-plan using a series of prompts.
 In the \emph{Discovery Phase}, the LLM is prompted to identify data items relevant for the query, such as relevant columns and datasets. In the \emph{Planning Phase}, the LLM is prompted to construct a sequence of steps to satisfy the user request (Logical Plan). The final \emph{Mapping Phase} is interleaved with \emph{Execution}: a physical operator is chosen for each of the logical steps and executed incrementally. 
 Once an operator is executed, we feed the results of the previous operator back to the LLM to choose the next operator (right). That allows the LLM to take the output of previous executions into account when choosing the physical operator and operator arguments (e.g. selection conditions as depicted in the figure) which avoids that faulty plans are generated.
}
 \vspace{-3.5ex}
 \label{fig:overview}
\end{figure*}

\noindent\textbf{Trend 1: Multi-Modal Data.}
In today's industries, huge amounts of non-relational multi-modal data (e.g. images, documents, sensor data, ...) need to be stored and processed, which has led to solutions that store data outside classical databases.
For instance, medical clinics need to store and analyze MRI scans and patient reports along with structured patient metadata.
Since these data modalities cannot be stored and queried easily in today's databases, they are usually stored in data lakes, which, however, makes gaining insights from these modalities hard.
To gain insights from such multi-modal data lakes, the data needs to be made accessible first, usually by manually constructing complex data processing pipelines.

While recently several pioneering systems have been proposed to ease the querying of multi-modal data in data lakes \cite{thorneNaturalLanguageProcessing2021, chenSymphonyNaturalLanguage2023, joDemonstrationThalamusDBAnswering2023},
these systems come with significant limitations.
For instance, the queries supported by these systems are often limited in complexity (e.g. only simple queries with a single value as answer are supported \cite{chenSymphonyNaturalLanguage2023}, or non-relational modalities are only used as filters \cite{joDemonstrationThalamusDBAnswering2023}), or they are limited to only a very few modalities \cite{thorneNaturalLanguageProcessing2021}.

An ideal data system for multi-modal data, instead, would allow all types of queries and automatically construct the data processing pipelines, that have previously been constructed manually.
This would allow users to formulate queries that combine information across modalities, and let them gain insights in a few minutes, for which they previously would have needed several days or weeks.

\noindent\textbf{Trend 2: Natural Language Interfaces.}
SQL, with its declarative nature, has initially been designed to be understandable by laypersons.
In practice, however, formulating complex queries in SQL requires profound knowledge of the language, making databases inaccessible to domain experts and management staff.
Usually, these persons are required to interact with specifically trained data scientists to obtain insights from data stored in databases, a process that often requires many iterations.

Hence, in recent years, Natural Language Interfaces for databases have emerged, which would allow laypersons to query databases using natural language.
However, existing approaches typically translate natural language into SQL \cite{yinTaBERTPretrainingJoint2020}
and are therefore limited to relational data.
Another direction are question-answering systems which work on modalities beyond tables \cite{chenSymphonyNaturalLanguage2023}.
However, question-answering systems only support queries that are much less expressive than what can be done with SQL.

\noindent\textbf{Our Vision.}
In this paper, we thus present a vision of a data system that can be used by laypersons using natural language and can query arbitrary data modalities while enabling complex user queries way beyond classical SQL as shown in Figure \ref{fig:teaser}.
In the example, a user queries a data lake of a museum that stores both metadata (stored as a table) and pictures of artworks (stored as images) exhibited in the museum.
To support such queries on multi-modal data, the natural language query must be translated into a complex processing pipeline that contains processing steps that can deal with multi-modal data.
For instance, in Figure \ref{fig:teaser}, the query is translated into a query plan that contains a \emph{Python} operator that can run arbitrary Python code, and a \emph{VisualQA} operator that extracts structured information from images.
In particular, in the example, the \emph{Python} operator extracts the century from a metadata column that stores the inception dates as strings and the \emph{VisualQA} operator is used to select all pictures that depict \emph{Madonna and Child}.
An important aspect is that the result for user queries in our system can range from single values, over tables, to even a plot.
Traditional query planners are clearly not able to come up with such a multi-modal query plan from a natural language query, since doing so requires \textbf{non-trivial reasoning over the user's intents, the available multi-modal data, as well as the effects of applying non-relational operators} to the data.

\section{Language Models as Query Planners}

In order to enable our vision, we propose \method{}, a novel query planner that leverages Large Language Models (LLMs) for compiling complex natural language queries over multi-modal data in executable plans.
Recently, LLMs such as GPT-4 \cite{openaiGPT4TechnicalReport2023}
have shown impressive results on various Natural Language Processing tasks such as translation, question answering, and reading comprehension.
Most important in our context, they have also shown to possess impressive \emph{reasoning capabilities} \cite{yaoReActSynergizingReasoning2023}.
In fact, it has already been shown that LLMs 
can reason not only about the content of multi-modal data \cite{zhangMultimodalChainofThoughtReasoning2023}
but also about the effect of applying certain tools on the data sources \cite{yangMMREACTPromptingChatGPT2023},
which enables e.g. a chatbot that can analyze an image that the user uploads \cite{wuVisualChatGPTTalking2023}.
However, to the best of our knowledge, no system has been proposed yet that operationalizes the reasoning capabilities of LLMs to construct complex query plans for multi-modal data lakes from natural language queries.

\subsection{Multi-phase Query Planning}

While LLMs provide reasoning capabilities as discussed before, it is still non-trivial to build a query planner that maps natural language user intents to executable query plans.
As shown in Figure \ref{fig:overview}, in \method{} we make use of a new multi-phase compilation strategy, which leverages carefully designed \emph{prompts} that contain all the necessary information about (1) the multi-modal data sources, (2) the available operators and (3) the query, which allows the LLM to come up with a query plan.
A major benefit of using prompts for query planning is that it is easy to plug in new operators (e.g. to process more modalities), as long as we provide all necessary information about their behavior in the prompt.
Moreover, our query planner based on LLMs is composed of \emph{multiple phases} which are based on the intuition of the phases of traditional query planning: first, by using a first prompt, the query planner maps the user query to a logical plan,  containing a high-level step-by-step (textual) description of what needs to be done.
Afterwards, using a separate prompt, each step is mapped to a concrete operator to obtain a physical plan that can be executed.
As we will see in our experiments, \method{} is thus indeed able to "reason" over user queries, data and available operators and can thus translate user queries into correct multi-modal query plans.

\subsection{Challenges}
Despite the capabilities of LLMs, there are many challenges when using LLMs for query planning. 
In the following, we discuss some of the main open research challenges for which we propose some ideas on how to tackle them in the following sections.

\noindent\textbf{Plan Executability.}
There are many causes why LLM-generated query plans might not be executable.
For instance, the LLM might provide the wrong inputs to an operator (e.g. a collection of images as input for a traditional SQL selection).
In these cases, plan execution will ``crash'' before the desired result can be computed.
One option to fix a non-executable plan (which is integrated into \method{}) is to use the LLM itself to fix the error by providing alternative plans as we discuss in Section \ref{sec:errors}.
While we show that LLMs are thus often able to fix errors this way, this is by far not enough to guarantee that query plans are executable as expected by users.

\noindent\textbf{Plan Correctness.}
Even if a query plan produced by an LLM is executable, there might still be ``logical flaws'' in the plan, which can lead to wrong query results.
For instance, important steps (e.g., a join) might be missing.
This is especially challenging since there is no feedback from the LLM on whether an executed plan is correct or not.
One option is to let users inspect the final plan and let them decide whether they trust it or whether they would like to improve it.
However, judging the correctness of such query plans can be difficult for laypersons.
Another idea is to improve the reasoning capabilities of LLMs by fine-tuning them to avoid typical reasoning errors. 
See Section \ref{sec:outlook} for a discussion on this direction.

\noindent\textbf{Plan Optimization.}
Finally, another important issue is that the plan generated by an LLM is not optimized, which is a problem since running non-optimized plans can result in huge runtime overheads.
However, optimizing multi-modal query plans requires reasoning over the runtime of complex multi-modal operators, such as \emph{VisualQA} or \emph{Python}, which is non-trivial.
One important component for optimizing multi-modal plans would be a learned cost model that captures the behavior of the multi-modal operators.
We discuss some further ideas on query optimization for LLM-based query planning in Section \ref{sec:outlook}.

\begin{figure}[t]
 \centering
 \includegraphics[width=1.0\linewidth]{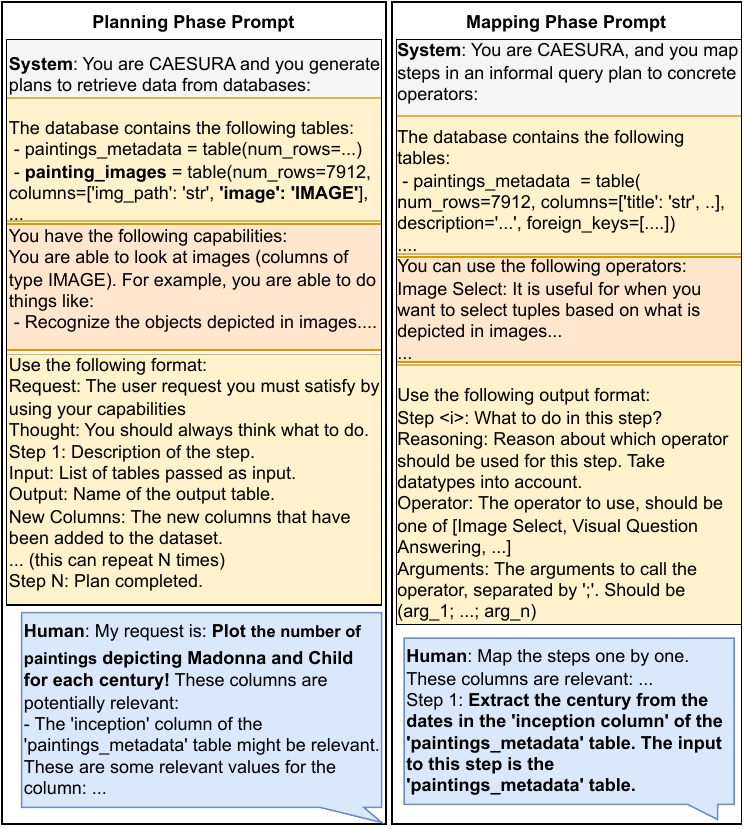}
 \vspace{-5.5ex}
 \caption{Example prompts for the Planning and Mapping Phase. Each prompt consists of two messages and contains all relevant information, e.g. data descriptions, operator/capability descriptions, etc. as well as an instruction telling the LLM what to do. In the planning phase, we additionally utilize in-context learning and provide a few example translations from query to logical plan for different domains at the very beginning of the prompt (not depicted).}
 \vspace{-5.5ex}
 \label{fig:prompt}
\end{figure}

\section{Overview of \method{}} \label{sec:overview}

In essence, as discussed before, in \method{} we orientate ourselves on the phases of traditional query planning and first generate a logical plan, which is afterwards translated to a physical plan.
However, in contrast to logical plans in databases, logical plans of \method{} consist of a description (in natural language) of the individual steps.
An example of such a logical plan is shown in Figure \ref{fig:overview} (below "Planning Phase").
Moreover, the physical plan contains operators that are very different from executable plans in databases.
An example physical plan is shown in the same Figure (below "Mapping Phase").

\subsection{Phases of Query Planning} \label{sec:phases}
Splitting the process of query planning into several phases allows us to tailor the prompts for query planning to the specific decisions of each phase.
See Figure \ref{fig:overview} for an overview of the three phases, which we elaborate in more detail in the following.
In a nutshell, we first identify the relevant data sources, then in the planning phase we let the LLM generate the logical plan, and finally, in the mapping phase we let the LLM select the operators to obtain a physical plan.

\noindent\textbf{Discovery Phase.}
In the first phase, we decide which data sources (e.g., in a data lake) provide relevant information for the current query.
We only briefly describe this phase, because the focus of this paper is on query planning.
In essence, \method{} first narrows down the relevant tables, image collections, etc. using dense retrieval (similar to Symphony \cite{chenSymphonyNaturalLanguage2023}). Afterwards, for tabular data sources, we prompt the LLM to decide which columns of the retrieved data are relevant to the user query.
The identified relevant data items are used to construct prompts for the next phases, e.g., to present the LLM with some relevant example values that help it to generate correct selection conditions.

\noindent\textbf{Planning Phase.}
In the planning phase, which is at the core of \method{}, the LLM is prompted to come up with a logical query plan that contains a natural language description of all steps necessary to satisfy the user's request.
Figure \ref{fig:prompt} (left) shows an example prompt for this.
The prompt consists of several parts: (1) a description of the data, (2) the capabilities of \method{}, (3) an output format description, and (4) finally the user query and an instruction telling the model to come up with a plan.
Notice how the multi-modal data is presented to the LLM: it is modeled as a special two-columned table where one column has the special datatype IMAGE.
The capabilities of \method{} describe the logical actions that \method{} can take with the help of the available operators as can be seen in the example.
Using this prompt, the LLM generates a stepwise (textual) plan which describes the logical plan in the output format specified in the input prompt.
The generated stepwise plan is then parsed by \method{} into a logical plan.
Moreover, in order to improve the quality of plans, we add a few examples of correct logical plans using few-shot prompting (not shown in the prompt in Figure \ref{fig:prompt}). This helps to instruct the model to produce plans in the desired output format.

\noindent\textbf{Mapping and Interleaved Execution.}
In the last phase, each previously determined logical step is mapped to a physical operator (and its input arguments) using a prompt similar to the one in Figure \ref{fig:prompt} (right).
The prompt for this phase contains a short summary of the operators and what they can be used for, which is inspired by recent work where LLM agents can use external tools \cite{wuVisualChatGPTTalking2023}.
Moreover, different from traditional query planning, we do not decide on all the physical operators for all logical steps at once.
Instead, we incrementally decide for each step and then execute it directly. For example, as shown in Figure \ref{fig:overview}, we first execute the \emph{VisualQA} operator before we decide to use a SQL selection for the next step.
This allows \method{} to react on the results returned by previous operations, which leads to more plans that are in fact executable.
For instance, in Figure \ref{fig:overview} after executing the \emph{VisualQA} operator, the LLM is able to construct a correct selection condition for the next step in the plan, based on the resulting values ("yes" and "no").

\subsection{Error Handling} \label{sec:errors}
With \method{} several errors can occur during query planning. 
In a nutshell, to deal with errors in \method{} , we use the LLM for error handling by adding the error message to the prompt and asking the LLM to fix the error.
However, this comes with the challenge that the root cause of an error is not known.
In particular, the root cause can also lie in any previously executed phase.
For instance, in the planning phase, the LLM could have decided to filter by a non-existent column, but the mistake is only noticed after choosing an operator and executing the step.

To fix such errors, we thus use the LLM to identify in which phase the error occurred, backtrack to it, fix the error, and rerun the subsequent phases.
For this purpose, we use an additional prompt containing a set of questions that encourage the LLM to reason about the error such as:
(1) What are the potential causes of this error?
(2) Explain in detail how this error could be fixed.
(3) Is there a flaw in my plan (Yes/No)?
(4) Is there a more suitable alternative plan (Yes/No)?
(5) Should a different tool be selected for any step (Yes/No)?
(6) Do the input arguments of some of the steps need to be updated (Yes/No)?
Parsing the responses to questions (3) + (4) allows us to determine whether to backtrack to the planning phase or if the mistake happened during the mapping phase.
To finally fix the error, the ideas from questions (1) + (2) and the original error message are added to the prompt to which we backtracked to, before it (and potentially subsequent phases) is executed again.
While this procedure allows \method{} to fix non-executable plans in many cases, it clearly does not guarantee plan executability or even plan correctness.
See Section \ref{sec:outlook} for further ideas on these issues. \begin{figure*}[t]
 \centering
 \includegraphics[width=0.85\linewidth]{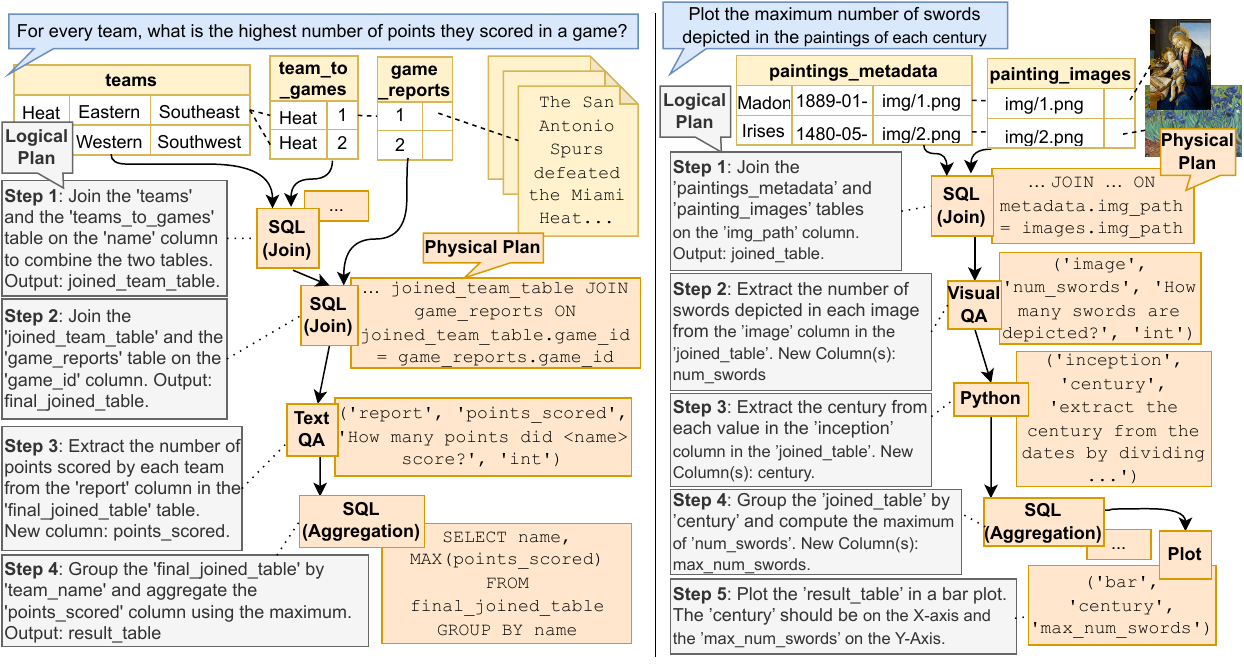}
 \vspace{-3.5ex}
 \caption{\method{} using GPT-4 is able to correctly translate the user queries to multi-modal query plans that contain TextQA, VisualQA and Python operators. The final physical plan, including input arguments for the operators, is shown in orange for both queries. For each operator, we also show the corresponding step of the GPT-4 generated logical plan (in grey).
 \method{} presents image and text collections as special tables (\emph{game\_reports} with columns \emph{game\_id} and \emph{report}; \emph{painting\_images} with columns \emph{img\_path} and \emph{image}) to the LLM, s.t. they can be the input to a regular join.
 The plans are presented as-is and are not optimized.
 The TextQA operator takes a question template as input, which is translated to questions by inserting different team names from the values in the table. The Python operator takes a description as input, which is translated to code using GPT-4.}
 \vspace{-2.5ex}
 \label{fig:anecdote}
\end{figure*}

\section{Initial Results} \label{sec:experiments}
In our experiments, we are primarily interested in whether \method{} is able to construct correct query plans.
Our prototype of \method{} has access to four multi-modal operators: (1) \emph{VisualQA} based on \emph{BLIP-2} \cite{liBLIP2BootstrappingLanguageImage2023}, (2) \emph{TextQA} based on \emph{BART} \cite{lewisBARTDenoisingSequencetoSequence2019}, (3) \emph{Python UDFs}, and (4) \emph{Image Select}, which selects images based on a description and is also based on BLIP-2.
It also has access to all \emph{relational operators} supported by SQLite and a plotting operator based on \emph{seaborn} \cite{Waskom2021}.

\noindent\textbf{Datasets.} Since there does not yet exist a benchmark for the scenarios we envision for \method{}, we constructed two multi-modal datasets.
(1) The \emph{artwork} dataset (with tables and images) resembles the example from Figures \ref{fig:teaser} and \ref{fig:overview}. The dataset contains a table about painting metadata as well as an image collection containing images of the artworks.
We use Wikidata to construct both the metadata table as well as the image corpus: for the metadata table, we extract title, inception, movement, etc. for all Wikidata entities that are instances of 'painting'.
(2) The second dataset is the \emph{rotowire} dataset (with tables and text) \cite{wisemanChallengesDatatoDocumentGeneration2017} which consists of textual game reports of basketball games, containing important statistics (e.g. the number of scored points) of players and teams that participated in each game.
We extend the textual reports by two tables for players and teams constructed from Wikidata.
These contain general information, such as name, conference, division, etc. for every team, and name, height, nationality, etc. for every player.

\subsection{Anecdotes}
Before we measure the system on a broad set of queries, we first highlight a set of correctly translated queries to illustrate that using LLMs for query planning is indeed promising.
We present two query plans obtained by translating two queries using \method{}, one for each dataset.
We use GPT-4 \cite{openaiGPT4TechnicalReport2023} as LLM for this experiment.

\noindent\textbf{Query 1 (on rotowire): For every team, what is the highest number of points they scored in a game?}
This query involves a join, usage of the TextQA operator as well as an aggregation. Figure \ref{fig:anecdote} (left) shows that \method{} was able to come up with a correct logical plan, that it then correctly translates into a physical plan.
The input arguments are chosen correctly as well.
Perhaps most impressive is that \method{} correctly uses the TextQA operator, which takes question templates as inputs.
During execution, these templates are instantiated by the operator using the values from the input table to generate questions like "How many points did Heat score?", which it then answers for all reports.
This allows the operator to separately extract the points scored by each team.

\noindent\textbf{Query 2 (on artwork): Plot the maximum number of swords depicted on the paintings of each century.}
This second and more complex query on the artwork data requires the inspection of the images, as well as the visualization of the results in the end.
Again, \method{} is able to correctly translate this query using two multi-modal operators: Python and VisualQA followed by a plot operator, as can be seen in Figure \ref{fig:anecdote} (right).

\subsection{Plan Quality}
Next, we evaluate \method{} on a larger set of queries, 24 for each dataset.
For this experiment, we are interested in the query planning abilities of LLMs.
Hence, we skip the data discovery step and assume perfect retrieval (to not measure retrieval performance).

The queries used in this experiment are clustered along several aspects.
In total, we have 16 queries asking for a single result value, 16 that ask for an output table, and 16 that ask for a plot.
Moreover, half of the queries require multi-modal data while the other half require only relational data.
Importantly, these queries were not used for tuning the prompts during the development. 
Table \ref{table:queries} shows the accuracies for the different query groups using ChatGPT-3.5 and GPT-4 as LLM.

We see that \method{} using GPT-4 is better than ChatGPT-3.5 and is even able to correctly translate 87.5\% of queries despite never being fine-tuned on the queries. 
The approach works especially well on the artwork dataset, where \method{} is able to translate all queries to correct query plans.
However, we also see that there is still room for improvement.
In particular, on the rotowire dataset, which consists of more tables and contains texts instead of images, only three-quarters of queries could be translated correctly.
In the next experiment, we analyze the mistakes \method{} makes when generating query plans.

\begin{table}
\small
\begin{center}
\begin{tabular}{ c c c c c}
  \toprule
    \textbf{Models} & \multicolumn{2}{c}{\textbf{ChatGPT-3.5}} & \multicolumn{2}{c}{\textbf{GPT-4}} \\
    Plan type & logical & physical & logical & physical \\
  \midrule
  \textbf{Artwork overall}  & 79.2\% & 70.8\% & 100\% & 100\% \\  
  \textbf{Rotowire overall} & 50.0\% & 41.7\% & 87.5\% & 75.0\% \\  
  \hline
  \textbf{Single modality} & 79.2\% & 75.0\% & 100\% & 92.7\% \\  
  \textbf{Multiple modalities} & 50.0\% & 37.5\% & 87.5\% & 83.3\% \\
  \hline
  \textbf{Single value} & 75.0\% & 62.5\% & 100\% & 93.8\% \\  
  \textbf{Table} & 68.8\% & 62.5\% & 87.5\% & 81.3\% \\ 
  \textbf{Plot} & 50.0\% & 43.8\% & 93.8\% & 87.5\% \\ 
  \hline
  \textbf{All} & 64.6\% & 56.2\% & 93.8\% & \textbf{87.5\%} \\
  \bottomrule
\end{tabular}
\caption{Correctly translated plans for the different datasets, modalities, and output formats. We show the percentage of correctly generated logical plans, as well as physical plans. }
\label{table:queries}
 \vspace{-8.5ex}
\end{center}
\end{table}

\subsection{Error Analysis}
In Table \ref{table:mistakes}, we categorize the errors and show the frequency of the errors for different categories.
We found that the possible errors are quite diverse: sometimes wrong physical operators were chosen or important steps were missing in the final plan (e.g. \method{} forgot to join).

The most common mistake for \method{} powered by GPT-4 was that it chose the wrong input arguments for physical operators (e.g., wrong parameters for SQL, a wrong question for QA, usage of non-existent column names).
For GPT-4, this happened for 3 out of the 48 queries.
For the smaller model, ChatGPT-3.5, we see that it had some more problems understanding the data correctly (see category \emph{Data Misunderstanding}), a mistake that happened only once with GPT-4.
In particular, ChatGPT-3.5 often tried to extract what is depicted in the image based on the title or the genre column of the metadata table.
Thus, it often avoided the usage of multi-modal operators and instead tried to solve everything using SQL, resulting in flawed plans.

Interestingly, there was one query on the rotowire dataset that both models could not translate at all:
\emph{How many games did each team lose?}
We speculate that this is the case because the query sounds simple and does not convey the operations necessary to answer it.
One possibility to answer this query would be to join the teams table and the game reports, use the TextQA operator to ask the question "Did <name> lose?", and then aggregate the losses.
Unfortunately, ChatGPT-3.5 tried to solve it using a single SQL query ignoring the text completely.
GPT-4, instead, was aware to use the text but it was not able to generate the correct operator for extracting the required information from the text.

\begin{table}
\small
\begin{center}
\begin{tabular}{ c c c c }
  \toprule
    \multicolumn{2}{c}{\textbf{Category}}  & ChatGPT-3.5 & GPT-4 \\
  \midrule
  \textbf{Impossible Actions} & logical & 4 & 2 \\
  \textbf{Data Misunderstanding} & logical & 9 & 1 \\
  \textbf{Illogical / Missing Steps} & logical & 3 & 0 \\
  \hline
  \textbf{Wrong Arguments} & physical & 3 & 3 \\
  \textbf{Wrong Tool} & physical & 1 & 0 \\

\bottomrule
\end{tabular}
\caption{Number of specific kinds of mistakes \method{} made during query planning. In the upper three categories the mistake occurred in the planning phase (i.e. wrong logical plan), and for the lower two the mistake occurred in the mapping phase. We see that the older model often does not understand the data correctly (e.g. it tries to determine what is depicted on a painting based on its title). }
\label{table:mistakes}
 \vspace{-8.5ex}
\end{center}
\end{table}
 \section{Research Directions} \label{sec:outlook}
We have seen that using today's state-of-the-art language models such as GPT-4 together with careful prompting yields promising results for multi-modal query planning.
However, there is still an abundance of interesting open challenges to be solved.
In particular, query planning is expected to yield correct and efficient plans, which cannot be guaranteed when LLMs are utilized.
In this Section, we explain our ideas on how these challenges could be overcome.

\noindent\textbf{Plan Executability and Correctness.}
While the reasoning capabilities of GPT-4 and similar LLMs are already impressive, they still make mistakes (see Table \ref{table:mistakes}).
In this regard, there are already first works that improve the reasoning capabilities of today's LLMs \cite{zelikmanSTaRBootstrappingReasoning2022}, and it remains to be seen if these improvements are translated to improved query planning.
Nevertheless, we speculate that it might not be enough to meet the strict quality constraints on query plans.
One interesting idea to push the data-reasoning capabilities of LLMs is to construct a fine-tuning dataset for query planning, similar to SPIDER \cite{yuSpiderLargeScaleHumanLabeled2019}.
SPIDER is a text-to-SQL dataset that boosted research on semantic parsing.
A similar fine-tuning dataset for multi-modal query planning could lead to comparable advancements in this field.
Such a fine-tuning dataset comes with several additional benefits on top of better reasoning skills and higher-quality plans.
Most importantly, it could be used to fine-tune smaller, open-source models, which resolve any privacy issues currently present when using external models via an API.
Moreover, the use of smaller LLMs would also reduce the computational (and monetary) cost of \method{}. 

\noindent\textbf{Plan Optimization.} \label{sec:plan-optimization}
In practice, there is also the need to generate runtime efficient query plans.
However, optimizing the resulting multi-modal query plan is far from trivial, since it requires reasoning over the runtime behavior of multi-modal operators.
This behavior can be hard to predict.
For instance, the execution of a Python operator can lead to vastly different runtimes depending on the Python code that is executed by the operator.
While there are already systems that are able to (partially) optimize UDFs in SQL queries \cite{rheinlanderOptimizationComplexDataflows2018}, they usually do not consider multi-modal data and the use of powerful Machine Learning models.
We believe an important step towards optimizing such multi-modal query plans is to learn cost models for multi-modal operators.
There is already a rich line of work for learned cost models (e.g. \cite{hilprechtZeroshotCostModels2022}).
However, so far these only capture the cost for traditional database operators and not the complex operators we consider in this paper.

\noindent\textbf{Security.}
Since we only have limited control over what is generated by an LLM, the LLM could theoretically generate malicious or destructive code to be executed over our data.
In our current prototype, we therefore limit e.g. generated SQL code to only SELECT statements and prevent running UPDATE, INSERT or DELETE statements that could maliciously manipulate data.
While we did not observe such behavior during the experiments, a more extensive analysis is necessary to rule out such concerns.

 \section{The Road Ahead} \label{sec:conclusion}
Gaining insights from multi-modal data is a difficult endeavor because it usually involves the manual creation of complex processing pipelines.
Hence, we present \method{}, a Language-Model-driven query planner that generates complex processing pipelines automatically from queries in natural language.
While \method{} shows first promising results, there are still a plethora of open challenges as discussed before.
In particular, today's language models suffer from reasoning difficulties and hallucinations, leading to query plans that crash or return wrong results, as well as plans that are sub-optimal in terms of runtime.
 
\begin{acks}
This research is funded by the Hochtief project \emph{AICO} (AI in Construction), by the BMBF and the state of Hesse as part of the NHR Program, as well as the HMWK cluster project \emph{3AI} (The Third Wave of AI). Finally, we want to thank hessian.AI at TU Darmstadt as well as DFKI Darmstadt.
\end{acks}

\bibliographystyle{ACM-Reference-Format}
\bibliography{bib}

\end{document}